\documentclass[prl,twocolumn,showpacs,letterpaper,showpacs,superscriptaddress,reprint]{revtex4}
\usepackage{graphicx,amsmath,amssymb,amsfonts,latexsym,color,dcolumn,bm,epsfig,subfigure}

\def\vcp#1{U_L(#1,z_0)}

\newcommand{\be}{\begin{equation}}
\newcommand{\ee}{\end{equation}}
\newcommand{\beqn}{\begin{eqnarray}}
\newcommand{\eeqn}{\end{eqnarray}}
\begin{document}

\title{Disorder in quantum vacuum: Casimir-induced localization of matter waves}

\author{G. A. Moreno}
\affiliation{IFIBA-Depto. de F\'{i}sica, FCEyN, UBA, Ciudad
Universitaria, 1428 Buenos Aires, Argentina}
\affiliation{Theoretical Division, MS B213, Los Alamos National Laboratory, Los Alamos, NM 87545, USA}

\author{R. Messina}
\affiliation{Laboratoire Kastler Brossel, case 74,
CNRS, ENS, UPMC, Campus Jussieu, F-75252 Paris Cedex 05, France}
\affiliation{SYRTE - Observatoire de Paris
61, avenue de l'Observatoire, F-75014 Paris, France}

\author{D. A. R. Dalvit}
\affiliation{Theoretical Division, MS B213, Los Alamos National Laboratory, Los Alamos, NM 87545, USA}

\author{A. Lambrecht}
\affiliation{Laboratoire Kastler Brossel, case 74,
CNRS, ENS, UPMC, Campus Jussieu, F-75252 Paris Cedex 05, France}

\author{P. A. Maia Neto}
\affiliation{Instituto de F\'{\i}sica, UFRJ, CP 68528, Rio de Janeiro,  RJ, 21941-972, Brazil}

\author{S. Reynaud}
\affiliation{Laboratoire Kastler Brossel, case 74,
CNRS, ENS, UPMC, Campus Jussieu, F-75252 Paris Cedex 05, France}

\date{\today}

\begin{abstract}
Disordered geometrical boundaries such as rough surfaces induce important modifications to the mode spectrum of the electromagnetic
quantum vacuum. 
In analogy to Anderson localization of waves induced by a random potential, here we show that the Casimir-Polder interaction between a cold atomic sample and a rough surface also produces localization phenomena. These effects, that represent a macroscopic manifestation
of disorder in quantum vacuum, should be observable with Bose-Einstein condensates expanding in proximity of rough surfaces.
\end{abstract}


\maketitle

%%%%%%%%%%%%%%%%%%%%%%%%%%%%%%%%%%%%%%%%%%%%%%%%%%%%%%%%%%%%%%%%%%%%%%%
{\it Introduction--}
Waves propagating in disordered potentials undergo multiple scattering processes that strongly affect their usual
diffusive transport and can result in localized states. In one dimensional systems Anderson localization is a ubiquitous phenomenon  \cite{Anderson}.
Recently it has been observed in a 1D Bose-Einstein condensate (BEC) expanding in the presence of a
random laser speckle field \cite{Billy} or a bi-chromatic potential \cite{Roati}. The asymptotic density profile shows exponential localization even in the weak disorder
limit  \cite{Sanchez-palencia}.  In this Letter we show that the high sensitivity to disorder of 1D cold atomic systems can be strong enough to yield localization of matter-waves due to disorder in vacuum.

The Casimir-Polder (CP) interaction between a BEC and a flat surface has been recently measured (see \cite{Obrecht} and references therein). Cold atoms act as local probes of the electromagnetic quantum vacuum, and they have been proposed as good probes of the influence of non-trivial geometrical effects on the CP atom-surface interaction potential \cite{Dalvit2008,MDC}. Charge and current quantum fluctuations in a rough surface induce a disordered CP potential that directly affects the dynamics of a BEC trapped close to such a surface, and may lead to localization of matter waves.
In this Letter we will investigate the localization properties of the BEC, which reveal the CP disordered interaction with a random surface.
An essential tool for computing the Casimir-Polder potential above a disordered surface is the scattering approach \cite{Lambrecht06}, that computes the Casimir interaction between bodies as a non-trivial multi-scattering process.

%%%%%%%%%%%%%%%%%%%%%%%%%%%%%%%%%%%%%%%%%%%%%%%%%%%%%%%%%%%%%%%%

{\it Atom-surface Casimir interaction-}
For simplicity, we will consider a surface with translational invariance along some direction (say $y$), and a generic uni-axial profile 
%\label{eq:h}
$h(x) = \sum_{i=1}^{\infty} h_i \cos(k_i x + \theta_i)$ (see Fig. 1).
Here $h_i$ are the amplitudes of the Fourier spectrum of the profile,  $\lambda_i=2 \pi/k_i$ the corresponding periods, and $\theta_i$ are
offsets. In the case of a corrugated surface these are fixed parameters, while for stochastic roughness they are random parameters
distributed according to certain probability distributions. In the following we will consider the latter case, with flat and independent probability distributions
in certain intervals (which is the simplest noise model to describe stochastic surfaces). 
A ground-state atom above such a surface is affected by a CP potential $U(x,z)$, where $z$ is the atom-surface distance (defined from a mean surface).
The exact computation of the CP potential for certain corrugated profiles,
such as 1D lamellar gratings \cite{Lambrecht2008,AMreyes}, can be performed via scattering theory. For stochastic roughness, however, exact results
do not exist. Previous works have computed the CP potential $U(x,z)$ by means of the pair-wise approximation \cite{Galina}. Here we calculate it using the scattering approach, that takes into account non-additivity effects. We expand the potential to second order in powers of $h_i$, that we assume are the smallest
length scales in the problem. In general, $U(x,z)$ has a part that depends only on $z$ (that for planar surfaces gives the usual CP force) and one that depends both on $x$ and $z$, that gives a lateral component to the CP force due to the lack of translational invariance. We shall denote this latter component as $U_L(x,z)$. The first order was already computed in \cite{Dalvit2008},
\begin{equation}
U^{(1)}_L(x,z) = - \frac{3 \hbar c \alpha(0)}{8 \pi^2 \epsilon_0 z^5} \sum_{i=1}^{\infty} h_i \, g^{(1)}(k_i z)\, \cos(k_i x + \theta_i),
\label{interaction}
\end{equation}
where $g^{(1)}$ is a response function that depends on the optical properties of the surface. Here we need the second-order correction
\cite{Riccardo}
\begin{eqnarray}\label{eq:second_order}
U_L^{(2)}(x,z)&=& - \frac{15\hbar c \alpha(0)}{32\pi^2\epsilon_0 z^6}  \sum_{i,j=1}^{\infty}  h_i h_j  \\
&\times&  [ \cos((k_i+k_j)x+\theta_i+\theta_j) g^{(2)} (k_i z,k_j z)  \nonumber  \\
&& + \cos((k_i-k_j)x+\theta_i-\theta_j)\, g^{(2)}(k_iz,-k_jz)]  , \nonumber
\end{eqnarray}
where the response function $g^{(2)}$ also depends on the material properties. In Fig. 1 we show $g^{(1)}$ and $g^{(2)}$ for a perfectly reflecting surface. The dimensionless kernel $g^{(1)}(kz)$ decays exponentially for large $kz$ (e.g., for a perfectly reflecting surface it is  $g^{(1)}({\cal Z})= e^{-{\cal Z}}
(1+{\cal Z} + 16 {\cal Z}^2/45 + {\cal Z}^3/45)$, with ${\cal Z}=kz$). The dimensionless kernel $g^{(2)}(k_i z, k_j z)$ typically decreases for large $z$ along generic directions in
the $k_i, k_j$ plane, but it grows along the $k_i=-k_j$ direction. However, since there is a maximum $k$ admissible in the
perturbative expansion (roughly given by $2\pi/h$, where $h$ is the typical order of magnitude of the height profile), this imposes a maximum
value to $k_i, k_j$ given by $2 \pi /h$, and therefore there exists an upper bound to the response function given by
$g^{(2)}(2\pi z/h,-2\pi z/h)$. Taking this fact into account, it follows that the second order perturbative expansion is sufficient for the purpose of this work \cite{Riccardo}.
%%%%%%%%%%%%%%%%%%%%%%%%%%%%%%%%%%%%%%%%%%%%%%%%%%%%%%%%%%%%%%%%%%%%%%
\begin{figure}
\begin{center}
\hspace{-20pt}
\scalebox{0.3}{\includegraphics{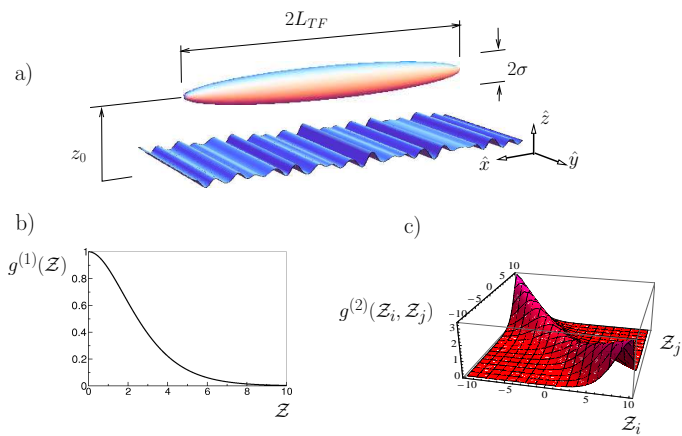}}
\vspace{-10pt}
\caption{a) Cigar-shaped BEC of width $\sigma$ and axial size $2L_{{\rm TF}}$ above rough surface. Here $L_{{\rm TF}}$ is the initial Thomas-Fermi length of the BEC. Dimensionless response functions for a perfectly reflecting surface: b)  $g^{(1)}({\cal Z})$ ( ${\cal Z}= k z$) and c) $g^{(2)}({\cal Z}_i,{\cal Z}_j)$ ( ${\cal Z}_i=k_iz$ and ${\cal Z}_j=k_jz$.)}
\label{fig:fig1}
\end{center}
\end{figure}
%%%%%%%%%%%%%%%%%%%%%%%%%%%%%%%%%%%%%%%%%%%%%%%%%%%%%%%%%%%%%%%%%%%%%%

%%%%%%%%%%%%%%%%%%%%%%%%%%%%%%%%%%%%%%%%%%%%%%%%%%%%%%%%%%%%%%%%%%%%%%
{\it Localization due to Casimir-Polder interactions-}
We consider the expansion of a tightly confined cigar-shaped BEC parallel to the rough uni-axial surface, so that its axis is perpendicular to the
corrugation lines (Fig. 1).
The effective 1D dynamics of the dilute BEC can be described by a mean-field wavefunction $\varphi(x,t)$ which evolves according to the 1D Gross-Pitaevskii (GP) equation \cite{pitaevskii}:
\begin{eqnarray}
\label{eq:fundamental}
i \hbar\, \partial_t \varphi(x,t) = - \frac{\hbar^2}{2m} \partial_x^2 \varphi(x,t) + \,U_L(x,z_0) \Theta(t)\,\varphi(x,t)\nonumber \\
+ \frac{m\omega_x^2x^2}{2} \Theta(-t)\,\varphi(x,t)+ g_{\rm eff} |\varphi(x,t)|^2 \varphi(x,t),
\end{eqnarray}
where $g_{\rm eff}=g/2 \pi \sigma^2$ is the effective coupling constant for the 1D problem ($\sigma$ is the width of the radial Gaussian profile). Although CP interactions are known to be non-additive, for dilute BECs additivity is a good approximation \cite{Antezza2004}.  
We assume that the harmonic potential
$m \omega_x^2x^2/2$ confines the system for $t<0$,  so that $\varphi(x,t)=e^{-i\mu t/\hbar} \varphi_0(x)$ with $\mu$ the chemical potential. At $t=0$ the axial trap  is turned off causing the BEC to expand in the presence of the disordered potential $U_L(x,z_0)$. Note that the distance to the surface, $z_0$, in Eq. (\ref{eq:fundamental}) is fixed during the evolution (the tight confinement freezes the radial motion), that is why all constant potential terms, including the $x$-independent component of the CP potential, are absorbed in a global phase-shift of the GP wavefunction. We intend to study the imprint left by CP forces on the asymptotic density profile $n(x,t)=|\varphi(x,t)|^2$ of the BEC after the expansion through the disordered potential.

More specifically we study the BEC density profile averaged over many realizations of $h(x)$, that we denote as
$\overline{n(x)}$. One of the parameters characterizing the evolution of the BEC is the root mean square $V_{\rm R}$ of the disordered potential, defined as $V_{\rm R}^2(z_0)=\overline{ (U_L(x,z_0)- \overline{U_L(x,z_0)})^2}$ ($x$ dependence disappears due to the random offsets $\theta_i$). 
When the strength of the disorder $V_{\rm R}(z_0)$ is much smaller than the chemical potential, the evolution can be treated analytically in perturbation theory \cite{Sanchez-palencia}. The applicability of such perturbative approach is valid in the CP context when relatively large distances $z_0$ are considered (note that $\lim_{z_0\to \infty} V_{\rm R}(z_0)=0$). However, when $V_{\rm R}(z_0)\approx \mu$ (small values of $z_0$), a full numerical approach is required.

\begin{figure}
\begin{center}
\hspace{-20pt}
\scalebox{0.4}{\includegraphics{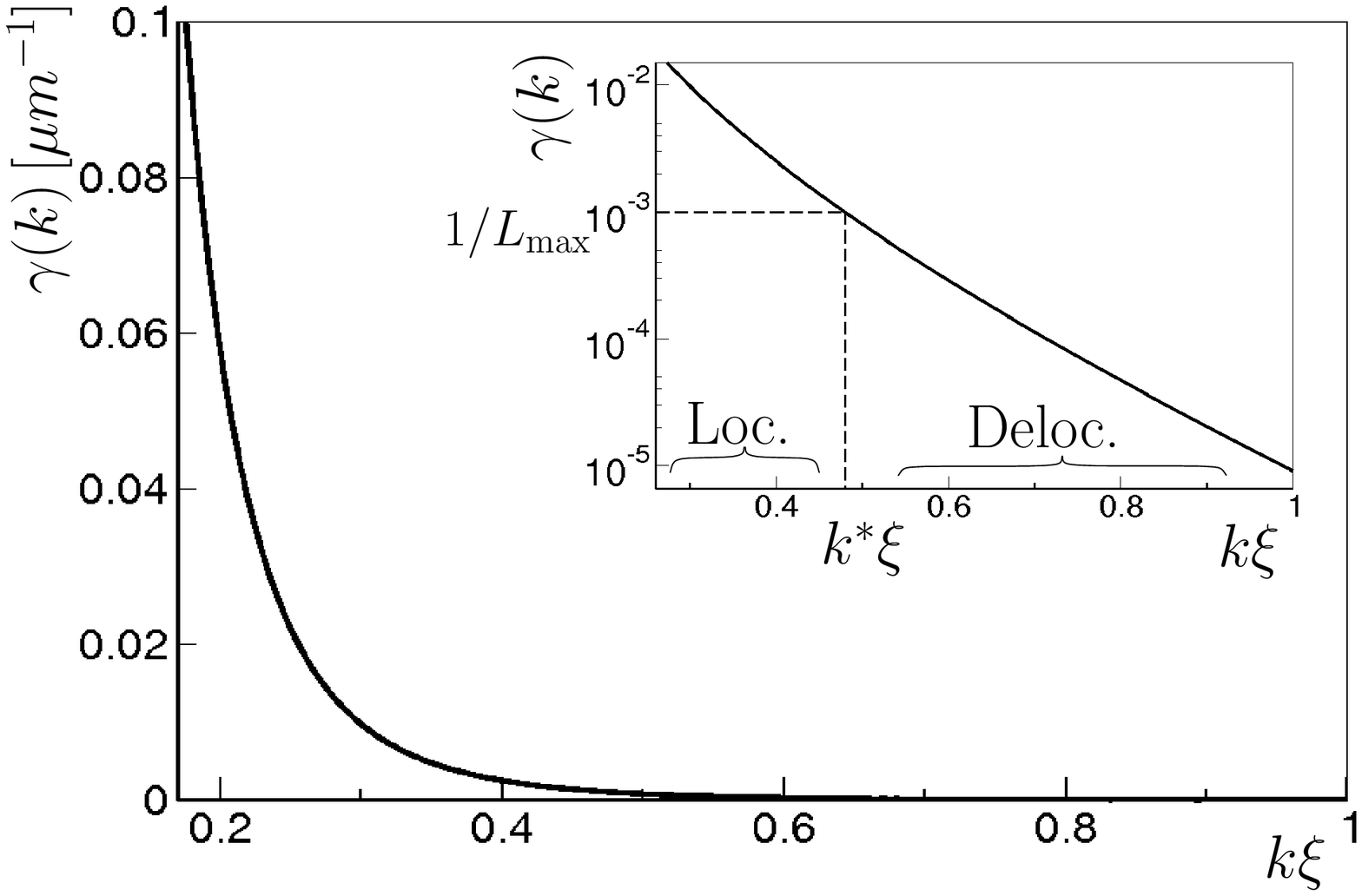}}
\vspace{-10pt}
\caption{$\gamma(k)$ vs. $k\xi$ from Eq.(\ref{eq:exponent}). 
The atom-surface distance is  $z_0=1.5\mu$m.
Inset: same data in Log-Lin scale, the maximum length scale to be measured $L_{\rm max}=1$mm, and the value $k^*=\gamma^{-1}(1/L_{\rm max})$ separating localized (Loc.) from delocalized (Deloc.) modes.}
\label{fig:fig2}
\end{center}
\end{figure}

Let us first consider the weak disorder case, $V_{\rm R}(z_0)\ll \mu$. When the trap is switched off the first stage of the expansion is driven by interaction term $g_{\rm eff} |\varphi(x,t)|^2 \varphi(x,t)$ in (\ref{eq:fundamental}) and the disordered potential can be neglected. After the initial expansion process the density becomes low enough to make the disordered term dominant and thus the system can be considered as a collection of non-interacting particles expanding in a weak disordered potential. In this situation the averaged density $\overline{n(x)}$ of localized atoms is fully characterized by the two-point correlator of the disordered CP potential \cite{Sanchez-palencia}: $
C(|x-x'|; z_0)=\overline{\vcp{x}\,\vcp{x'}\,}$. In the following we will omit writing the explicit dependence on the parameter $z_0$. This quantity determines to which extent a non-interacting particle with momentum $k$ localizes in the random potential. The information about the localization length is given by the inverse of the so-called Lyapunov exponent \cite{lifshits}, 
$\gamma(k)=(m/4\hbar^2E_k) \int_{-\infty}^\infty C(x) \cos(2kx) \,{\rm d}x$, 
where $E_k=\hbar^2 k^2 / 2m$. The consistency of this perturbative approach requires $\gamma(k)\ll k$ to hold, that is the typical length scale of the exponential localization must be larger than the typical wave-length of the unperturbed single-particle state. Deviations from a pure exponential behavior were calculated in \cite{gogolin}, and used in \cite{Sanchez-palencia}  to find an expression for the averaged density profile, namely:
\be
\label{eq:naverage}
\overline{n(x)}=\frac{3N\xi}{2} \int_0 ^{1/\xi} (1-k^2\xi^2) \, \overline{|\phi_k(x)|^2}\,{\rm d}k \,\, ,
\ee
where $\xi=\hbar/\sqrt{4m\mu}$ is the healing length of the BEC and the averaged modulus squared of the single particle wavefunction $\phi_k$ is
\beqn
\label{eq:gogolin}
\overline{| \phi_k (x) |^2} &=& \frac{\pi^2 \gamma(k)}{2} \int_0^\infty \,u\, {\rm sinh}(\pi u) \\
&\times&  \left( \frac{1+u^2}{1+{\rm cosh} (\pi u)} \right)^2 \; e^{- 2 (1+u^2) \gamma(k) |x|}\,{\rm d}u \nonumber .
\eeqn
Using these relations $\overline{n(x)}$ can be calculated once $C(x)$ is given. In general, two different asymptotic regimes for the wings of the density profile can be identified, depending on the healing length of the BEC, the spectral content of $C$, and the maximum length scale $L_{\rm max}$ that can be measured in the system (e.g. the spatial observation window of the measurement set up). When the above system is solved for $\overline{n(x)}$ assuming that $\gamma(k)\neq 0 \,\forall k \in (0,1/\xi)$ and $L_{\rm max}\to \infty$ an exponentially localized profile is always found. On the other hand, when
$\gamma(k)$ vanishes in an interval  $\tilde k \leq k \leq 1/\xi$ and $L_{\rm max}\to \infty$, there exists an effective sharp mobility edge at $k=\tilde k$, the modes with $k>\tilde k$ continue to expand and the localized modes with $k<\tilde k$ give rise to an algebraic-decay of the form $\overline{n(x)}\propto 1/x^\nu$. This is typically the case of the correlator of the laser speckle potential, that has a sharp cut-off. In contrast, the function $\gamma(k)$ for the CP disordered potential of the stochastic surface considered here reads: 
\begin{equation}
\gamma(k)=\frac{m\pi^2 F^2(z_0)}{2\hbar^2E_k(2k)^2} \, \overline{h(x)^2}\, P(\pi/k) \, \left(g^{(1)}(2k z_0)\right)^2 ,\label{eq:exponent}
\end{equation}
where $\overline{h(x)^2}$ is the mean of $h(x)^2$ at any point $x$, and $P(\lambda)$ is the probability density for each $\lambda_i$ (in this case flat) and $F(z_0)=3 \hbar c \alpha(0)/8 \pi^2 \epsilon_0 z_0^5$. Hence $\gamma$ is exponentially suppressed (see Fig. \ref{fig:fig2}).
In the CP case exponential localization is also found if $L_{\rm max}\to \infty$. However, for a finite value of $L_{\rm max}$, the situation may change. In fact, in typical cases where the perturbative approach is valid in this context the Lyapunov exponent satisfies
$\gamma(1/\xi)L_{\rm max}\ll 1$ making the modes with $k^* \gtrsim \gamma^{-1}(1/L_{\rm max})$ effectively delocalized, see the inset in Fig.
\ref{fig:fig2} (here $\gamma^{-1}$ denotes the inverse function of $\gamma$). Thus, in a restricted region of size $2L_{\rm max}$, the function $\overline{n(x)}$ does not decay exponentially but rather algebraically (this assertion is approximate and becomes exact in the limit $\gamma(1/\xi)L_{\rm max}\to 0$), the fast decay of $\gamma(k)$ results in an effective mobility edge at $k^*$. However, the transition between the two limiting regimes is not sharp in the CP context, and in principle one can pass continuously between both varying $L_{\rm max}$. Finally, note that this is an external parameter which does not affect the system but only has to do with the measurement process.

%%%%%%%%%%%%%%%%%%%%%%%%%%%%%%%%%%%%%%%%%%%%%%%%%%%%%%%%%%%%%%%%%%%%%%
%%%%%%%%%%%%%%%%%%%%%%%%%%%%%%%%%%%%%%%%%%%%%%%%%%%%%%%%%%%%%%%%%%%%%%

%%%%%%%%%%%%%%%%%%%%%%%%%%%%%%%%%%%%%%%%%%%%%%%%%%%%%%%%%%%%%%%%%%%%%%
\begin{figure}
\begin{center}
\hspace{-20pt}
\scalebox{0.4}{\includegraphics{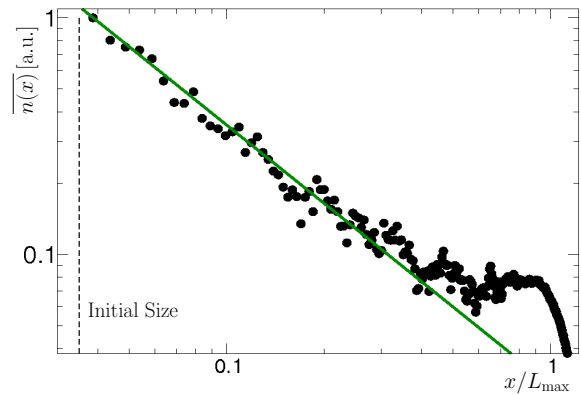}}
\vspace{-10pt}
\caption{BEC density (arbitrary units) vs. position. Both the perturbative theory described by
Eqs.(\ref{eq:naverage},\ref{eq:gogolin},\ref{eq:exponent}) (solid) and the full numerical simulation (dots) are computed using the first order approximation for the CP potential at $z_0=1.5\mu$m. The surface profile is averaged over $40$ realizations. Time corresponds to $\omega_x t=28$.}
\label{fig:fig3}
\end{center}
\end{figure}
%%%%%%%%%%%%%%%%%%%%%%%%%%%%%%%%%%%%%%%%%%%%%%%%%%%%%%%%%%%%%%%%%%%%%%

We now compare this perturbative approach with a full numerical simulation.  We consider a BEC of $N=10^2$ $^{87}$Rb atoms tightly confined in the radial direction. The width of the ground state is assumed to be  $\sigma=0.25\mu$m (i.e., radial trapping frequency
$\omega_r=2\pi \times 286$Hz) and its axial size $35\mu$m (i.e.,  $\omega_x=2\pi \times 2.75$Hz). For these values the healing length of the BEC is $\xi=0.85\mu$m. A perfectly conducting uni-axial rough surface is brought to close proximity of the BEC, with typical distances of $z_0 \approx 1\mu$m. The stochastic surfaces are generated using between $15$ and $25$ harmonics, and the parameters
$h_i$,  $\lambda_i$, and $\theta_i$ are taken as independent random variables with flat probability distributions satisfying
$h_i\in[0,200]$nm, $\theta_i\in[0,2\pi]$, $\lambda_i \in[\lambda_{\rm min},\lambda_{\rm max}]$. Note that the approximation we are using to evaluate the CP lateral potential assumes $\lambda_{\rm min}>h$, however this is not an important restriction because the exponential suppression of the modes with the factor $z_0/\lambda_{\rm min}$ makes the evolution insensitive to the lower limit. In Fig. 3 we compare a direct numerical simulation with the perturbative approach for the function $\gamma(k)$ shown in Fig. 2, which corresponds to a surface with $25$ harmonics and $\lambda_i \in[1,20]\mu$m at a distance of $z_0=1.5\mu$m of the BEC (in this situation the mean square root of the noisy potential is $V_{\rm R}=0.089\,\mu$). Even in this pertubative regime the integral in (\ref{eq:gogolin}) must be calculated numerically. In the preceding case we have taken into account only the first order correction to the lateral CP potential because it is a good approximation in such regime (this was verified in the numerical simulations). Note that, while the BEC evolves, the averaged density profile $\overline{n(x,t)}$ (here truncated after $40$ realizations of the surface profile) approaches to the asymptotic $t\to \infty$ solid line $\overline{n(x)}$ predicted by the theory \cite{Sanchez-palencia}. Because of the finite value of the time variable used here, $\omega_x t=28$, it can be seen that the edge at the right is expanding yet, snapshots at shorter times show a bigger elbow for smaller positions. The space window chosen corresponds to $L_{\rm max}=1$mm and, in that scale, a log-log graph makes the prediction of the perturbative approach to be a straight line (slope constant within $1\%$ accuracy).

%%%%%%%%%%%%%%%%%%%%%%%%%%%%%%%%%%%%%%%%%%%%%%%%%%%%%%%%%%%%%%%%%%%%%%%
\begin{figure}
\begin{center}
\hspace{-20pt}
\scalebox{0.4}{\includegraphics{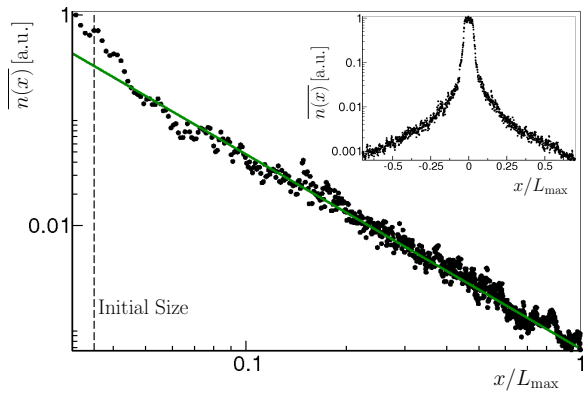}}
\vspace{-10pt}
\caption{BEC density (arbitrary units) vs. position, after $\omega_x t=14$. The profile is averaged over $40$ realizations at a distance of $z_0=1.0 \mu$m. Numerical simulation (dots) includes both the first and second order terms of the lateral CP potential $U_L(x,z)$. The wing is fitted by a power law $\overline{n(x)} \propto 1/x^\nu$ with $\nu = 1.84$ (solid). Inset: zoom of the numerical data in Log-Lin scale.}
\label{fig:fig4}
\end{center}
\end{figure}
%%%%%%%%%%%%%%%%%%%%%%%%%%%%%%%%%%%%%%%%%%%%%%%%%%%%%
In the following we consider stronger perturbations, so that $V_{\rm R}(z_0)$ is no longer much smaller than $\mu$. In such circumstances the previous perturbative treatment breaks down since the CP potential cannot be disregarded in the first stage of the expansion. Therefore a fully numerical method is required to solve Eq. (3). For distances of about $z_0 \approx 1 \mu$m the second order correction to the lateral CP potential Eq. (\ref{eq:second_order}) also becomes important, so in the full numerical simulation we solve exactly the evolution of the BEC taking into account these second order effects. The results are shown in Fig. 4, where we plot a typical stationary density profile averaged over $40$ realizations of the surface at $z_0=1\mu$m from a surface with $15$ harmonics in the range $\lambda_i \in [1,8]\mu$m and $h_i\in[0,200]$nm (for these parameters $V_{\rm R} = 0.5 \mu$). Most of the atoms are trapped in a high density core of typical size given by the initial size of the BEC due to single barrier reflections (similar to the strong disordered regime considered in \cite{Sanchez-palencia}), the atoms leaving the core develop a density profile with algebraic-like wings. Again the exponential suppression of the CP potential makes the evolution insensitive to modifications in the lower limit of the range of wavelengths $\lambda_{\rm min}$, the spatial cut off frequency being imposed by $z_0$ (this was also confirmed numerically pushing the surface spectrum to shorter wavelengths). 
There is a high sensitivity with the distance to the surface due to two effects, the strong power-law decay in $z_0$ of the CP potential, and the exponential suppression of the modes with $k z_0$. Their combined effect  makes the system pass rapidly from the perturbative to the non-perturbative regimes. In fact it can also be seen that for configurations such as the one presented here distances of about $z_0 \approx 3\mu$m make $V_R$ so small compared with the typical kinetic energy that the effect of the CP potential on the expansion of the BEC is negligible.
Finally, it is worth mentioning that the localization phenomena found for the parameters used in our work may break down for stronger BEC non-linearities $g_{\rm eff}$ in Eq. (\ref{eq:fundamental}) \cite{pikovsky}.

We have shown how quantum fluctuating interactions in non-trivial geometries affect transport properties of matter waves. We have found that a quasi-1D Bose-Einstein condensate expanding close to a rough surface undergoes algebraic localization. The perturbative prediction for the density profile of the BEC localized via the disordered Casimir-Polder interaction was shown to be in good agreement with a fully numerical approach. 

GAM thanks E. Calzetta  and J. J. Z\'arate for helpful discussions. This work was partially supported by CONICET, UBA, ANPCyT, Los Alamos LDRD program, CAPES-COFECUB, ESF Research Networking Programme CASIMIR (www.casimir-network.com), CNPq and FAPERJ-CNE.

%%%%%%%%%%%%

\end{document}